\documentclass[reprint,superscriptaddress,amsmath,amssymb,jcp]{revtex4-1}
\pdfoutput=1 
\usepackage{dcolumn}
\usepackage{epsfig}
\usepackage{graphics}
\usepackage[T1]{fontenc}
\usepackage{bm}
\usepackage{hyperref}
\usepackage{amssymb}
\newcommand{\ddst}{false}

\begin{document}

\title{Nature of Radiation-Induced Defects in Quartz}

\author{Bu Wang}
 \affiliation{Physics of AmoRphous and Inorganic Solids Laboratory (PARISlab), Department of Civil and Environmental Engineering,
University of California, Los Angeles, CA, USA}

\author{Yingtian Yu}
 \affiliation{Physics of AmoRphous and Inorganic Solids Laboratory (PARISlab), Department of Civil and Environmental Engineering,
University of California, Los Angeles, CA, USA}

\author{Isabella Pignatelli}
 \affiliation{Laboratory for the Chemistry of Construction Materials (LC$^2$),
Department of Civil and Environmental Engineering,
University of California, Los Angeles, CA, USA}

\author{Gaurav N. Sant}
 \affiliation{Laboratory for the Chemistry of Construction Materials (LC$^2$),
Department of Civil and Environmental Engineering,
University of California, Los Angeles, CA, USA}
 \affiliation{California Nanosystems Institute (CNSI), University of California, Los Angeles, CA, USA}

\author{M. Bauchy}
 \email[Contact: ]{bauchy@ucla.edu}
 \homepage[\\Homepage: ]{http://mathieu.bauchy.com}
 \affiliation{Physics of AmoRphous and Inorganic Solids Laboratory (PARISlab), Department of Civil and Environmental Engineering,
University of California, Los Angeles, CA, USA}

\date{\today}

\begin{abstract}
Although quartz ($\rm \alpha$-form) is a mineral used in numerous applications wherein radiation exposure is an issue, the nature of the atomistic defects formed during radiation-induced damage have not been fully clarified. Especially, the extent of oxygen vacancy formation is still debated, which is an issue of primary importance as optical techniques based on charged oxygen vacancies have been utilized to assess the level of radiation damage in quartz. In this paper, molecular dynamics (MD) simulations are applied to study the effects of ballistic impacts on the atomic network of quartz. We show that the defects that are formed mainly consist of over-coordinated Si and O, as well as Si--O connectivity defects, e.g., small Si--O rings and edge-sharing Si tetrahedra. Oxygen vacancies, on the contrary, are found in relatively low abundance, suggesting that characterizations based on $E^{\prime}$ centers do not adequately capture radiation-induced structural damage in quartz. Finally, we evaluate the dependence on the incident energy, of the amount of each type of the point defects formed, and quantify unambiguously the threshold displacement energies for both O and Si atoms. These results provide a comprehensive basis to assess the nature and extent of radiation damage in quartz.
\end{abstract}

\maketitle

\section{Introduction}
\label{sec:intro}

Quartz (the dominant crystalline form of SiO$_{2}$ under ambient conditions) is one of the most abundant minerals in Earth's crust. It is widely used in engineering applications, e.g., in construction materials, semiconductors, integrated optics, ultrasonic devices, etc. \cite{Fischer1983, Lell1968, Sim1991, Phillips2010, Manzano-Santamaría2012}. Many of these applications require the material to withstand exposure to radiations. For example, when used as a concrete aggregate, quartz can be subjected to neutron radiation in nuclear power plants \cite{Field2015,LePape2015}. Therefore, it is important to establish a good understanding of the nature of the damages induced by radiations, so that the material durability and reliability can be improved. In addition, improved knowledge of radiation-induced damage mechanisms in quartz could also inform radiation damage sensitivities in similar materials such as silicate glasses, which are also of substantial interest for radiation related applications including nuclear waste encasement and optics in fusion reactors \cite{Ibarra2004}.

Numerous experimental studies have been conducted to assess the nature and extent of damage induced in quartz by various types of irradiations, and often using neutrons and different types of ions \cite{Wittels1954,Simon1957,Primak1958,Hines1960,Fischer1983,Macaulay-Newcombe1984,Manzano-Santamaría2012}. It is generally admitted that different types of irradiation result in similar damage to the atomic structure \cite{Hines1960}. Such damage has been described in term of an accumulation of point defects, which eventually causes structural amorphization \cite{Primak1958,Fischer1983}. However, the final amorphized structure is found to be different from that of glassy silica, and the detailed nature of the formed point defects has not been clearly identified \cite{Simon1957,Primak1958}. Numerous studies have focused on the oxygen vacancies created during irradiation \cite{Arnold1965,Devine1983,Fischer1983,Devine1984}. Indeed, it is well-known that the oxygen vacancies present in quartz after ionization or electronic radiations can trap electrons, forming the so-called $E^{\prime}$ center, a color center that modifies the optical properties of quartz \cite{Weeks2008}. This has been used to assess, by optical techniques, the level of damage and the threshold displacement energy ($E_{\rm d}$) \cite{Arnold1965,Fischer1983}, a critical quantity commonly used to estimate radiation damage \cite{Ziegler1985}. However, the reported $E_{\rm d}$ is substantially larger for oxygen than for silicon. This appears counterintuitive just by considering that four bonds need to be broken to displace a silicon atom, but the rupture of only two of such bonds are needed for oxygen displacement \cite{Arnold1965}. The reported $E_{\rm d}$ for oxygen in quartz is also almost twice as large compared to that in fused silica \cite{Mota2005}, which is unexpected. Additional experimental evidence suggests that $E^{\prime}$ center creation is accompanied by substantial formation of other unidentified defects (up to 500 other defects for every $E^{\prime}$ center), suggesting that relying only on the oxygen vacancies to assess the damage and determine relevant properties may not be appropriate \cite{Devine1983,Devine1984}. These complexities in structural damage models impede better understanding of the damage process, and also create difficulties in accurately predicting the damage evolved in materials, for engineering purposes.

Molecular dynamics (MD) simulations provide a unique ability to access the damage process in irradiated quartz. In the past decade, several studies have applied MD to explicitly simulate radiation damage in silicates \cite{Delaye,Weber1997,Kubota2003,Mota2005,Mota2004,Yu2009,Delaye2011}. For crystalline quartz, however, such studies have been limited. To this end, herein, we perform MD simulations of radiation-induced atomic structural damage in quartz. By detailed analysis of MD trajectories and careful consideration of the creation and annihilation oxygen vacancies, we confirm that oxygen vacancies only account for a small percentage of all the induced defects. The nature and quantity of the dominant defects and their dependence on the incident energy are evaluated. This information, together with improved estimations of threshold displacement energies for both oxygen and silicon atoms, allow for more accurate assessment of the radiation damage in quartz. Finally, we also examine damage to the connectivity of the Si--O network, such as the modification of the ring size and the appearance of edge-sharing Si tetrahedra, which could serve as a structural basis for the radiation-induced amorphization at higher dosage.

\section{Simulation methods}
\label{sec:sim_meth}
Molecular dynamics simulations were carried out with the LAMMPS code \cite{Plimpton1995}. To ensure that both pristine and amorphized quartz can be properly simulated, we selected the BKS potential, \cite{Beest1990} a widely adopted potential set for silica. Note that this potential has been previously used to study radiation damage in glassy silica and has been shown to provide realistic predictions \cite{Yu2009,Delaye2011,Phillips2010}. The potential, although reliable at equilibrium separation, does not provide any physical description of interactions when the atoms reach closer distances with respect to each other, as observed during high-energy collisions. These interactions can, instead, be described with the ZBL potential, which has been designed to tackle such atomic collisions \cite{Ziegler1985}. For a smooth transition, the BKS and ZBL potentials are linked by a polynomial function. The first and second derivatives are examined carefully when adjusting the polynomial transition to ensure that no additional extrema (that would result in unrealistic metastable states) are created by the process. The cutoff for the BKS potentials is set to 5.5 \AA,\ following the previously reported protocol, \cite{Yuan2014,Wang2015} and the ZBL potentials take effect below 0.8 \AA\ for Si--O, 1 \AA\ for O--O, and 1 \AA\ for Si--Si interactions, respectively.

Radiation damage is simulated within a supercell of quartz. The deposition of irradiation energy is simulated such that an atom in the supercell is randomly selected and accelerated with a given kinetic energy, corresponding to the impact (incident) energy. It must be noticed that the absorption of irradiation energy is not explicitly simulated here. However, the kinetic energy used to accelerate the atom can be readily translated from the incident irradiation energy for ion implantation, or the incident radiation energy for neutron, electron or electrostatic radiations, if the absorption behavior is known. Therefore, for convenience, this acceleration energy will be referred to as the "incident energy" in the following sections.

Following the energy deposition, the kinetic atom leads to a cascade of collisions with the neighboring atoms. A spherical region is created around the region that would be potentially damaged. Within this region, the temperature of the atoms is not controlled. A thermostat is instead placed on the rest of the system, outside of the spherical region, and set to 300K. Hence, the thermostat does not influence the true dynamics of the displaced atoms during the ballistic cascade, whereas the rest of the system, which is not directly affected by the cascade, serves as heat sink, ensuring that local heating and quenching effects are properly reproduced. Note that, since the damage region is dynamically created following the irradiation energy deposition, the need to maintain a specific area for heat sink is eliminated, allowing damage to be initiated from any atom in the system.

During the initial cascade, a variable time step is invoked to avoid excessive displacements of atoms during a single step, with a maximum time step of 1 fs. Depending on the incident energy, most collisions occur within the first 1 to 3 ps, before the temperature of the system eventually stabilizes to 300K. Atomic trajectories during the first 3 ps are therefore analyzed in high-resolution to track the collisions. Afterwards, the simulation is continued for another 50 ps, allowing the structure to relax. To obtain meaningful statistical data, this process is repeated 100 times for each incident energy. At each cycle, an impact atom is randomly selected. Both the size of the supercell and the spherical region are selected based on the incident energy. For 300, 600 and 900 eV incident energies, 8$\times$9$\times$7 (4536 atoms), 10$\times$12$\times$9 (9720 atoms), 13$\times$15$\times$12 (21060 atoms) quartz supercells, and 16 \AA,\ 25 \AA,\ and 33 \AA\ radius spherical regions are used, respectively (see further details in the supporting information).

\section{Results and discussions}
\label{sec:res_dis}

\subsection{Structural damage}
\label{stru_dam}

Radiation damage in oxides is commonly described by point defect formation \cite{Weber1997,Weber1998}. For quartz too, this has been proposed as the primary damage manifestation at low dosage \cite{Primak1958,Fischer1983}. Because the oxygen vacancy in quartz can trap localized charge, forming the $E^{\prime}$ (color) center, such defects have been used to study damage in neutron radiated and ion-implanted quartz \cite{Arnold1965,Fischer1983}. However, we find that oxygen vacancy formation is not the main form of damage involved. 

\begin{figure*}
\begin{center}
\scalebox{0.5}{\includegraphics*[width=\linewidth, keepaspectratio=true, draft=\ddst]{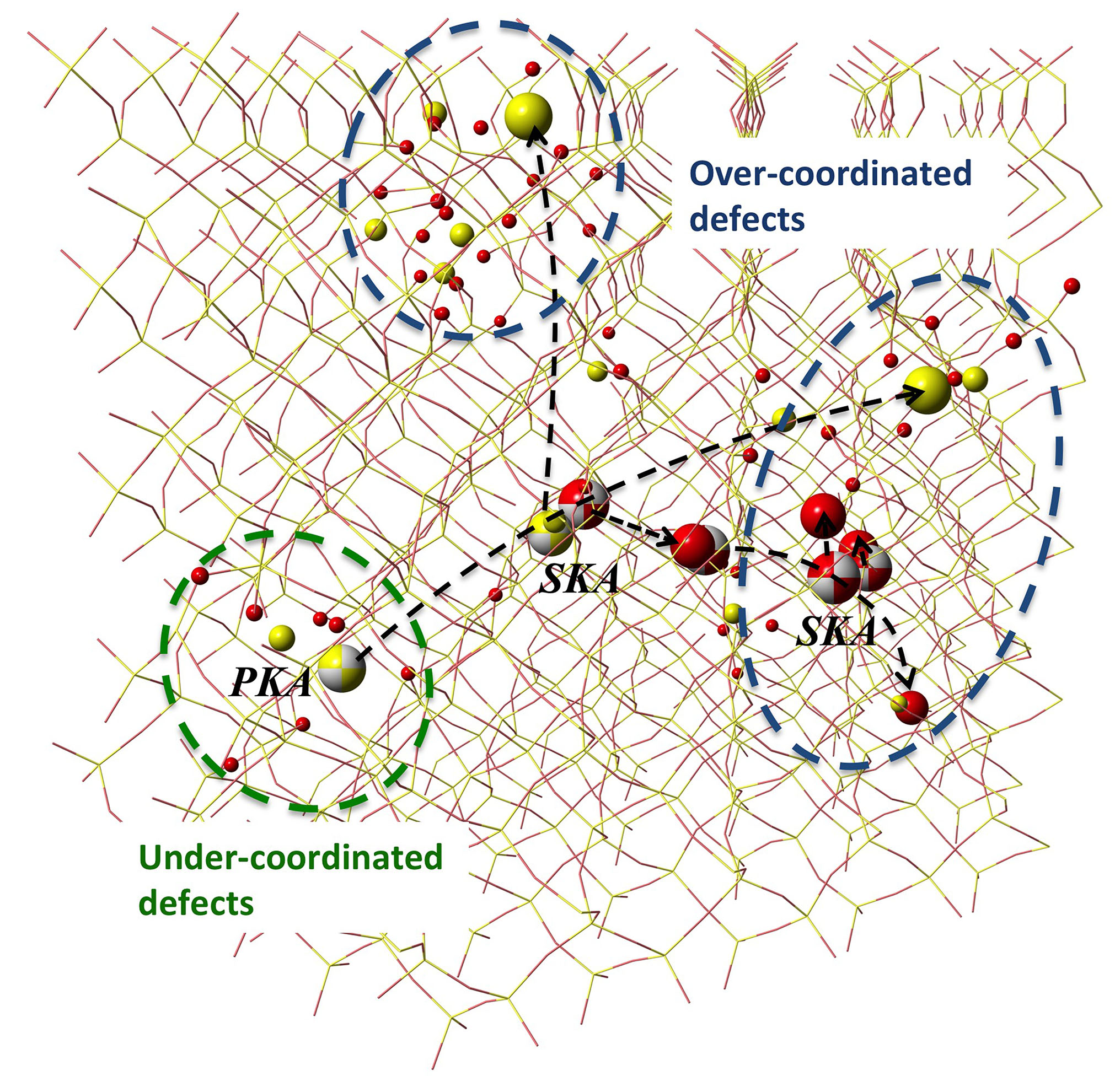}}
\caption{\label{fig:1} An example 600 eV cascade trajectory. Silicon and oxygen atoms with a displacement greater than 1 \AA\ are shown as yellow and red spheres, respectively. Major high-velocity atoms during the cascade, such as the primary knock-on atom (PKA) and secondary knock-on atom (SKA), are highlighted with exaggerated sizes, and their initial positions are shown with beach balls.}
\end{center}
\end{figure*}

\begin{figure*}
\begin{center}
\includegraphics*[width=\linewidth, keepaspectratio=true, draft=\ddst]{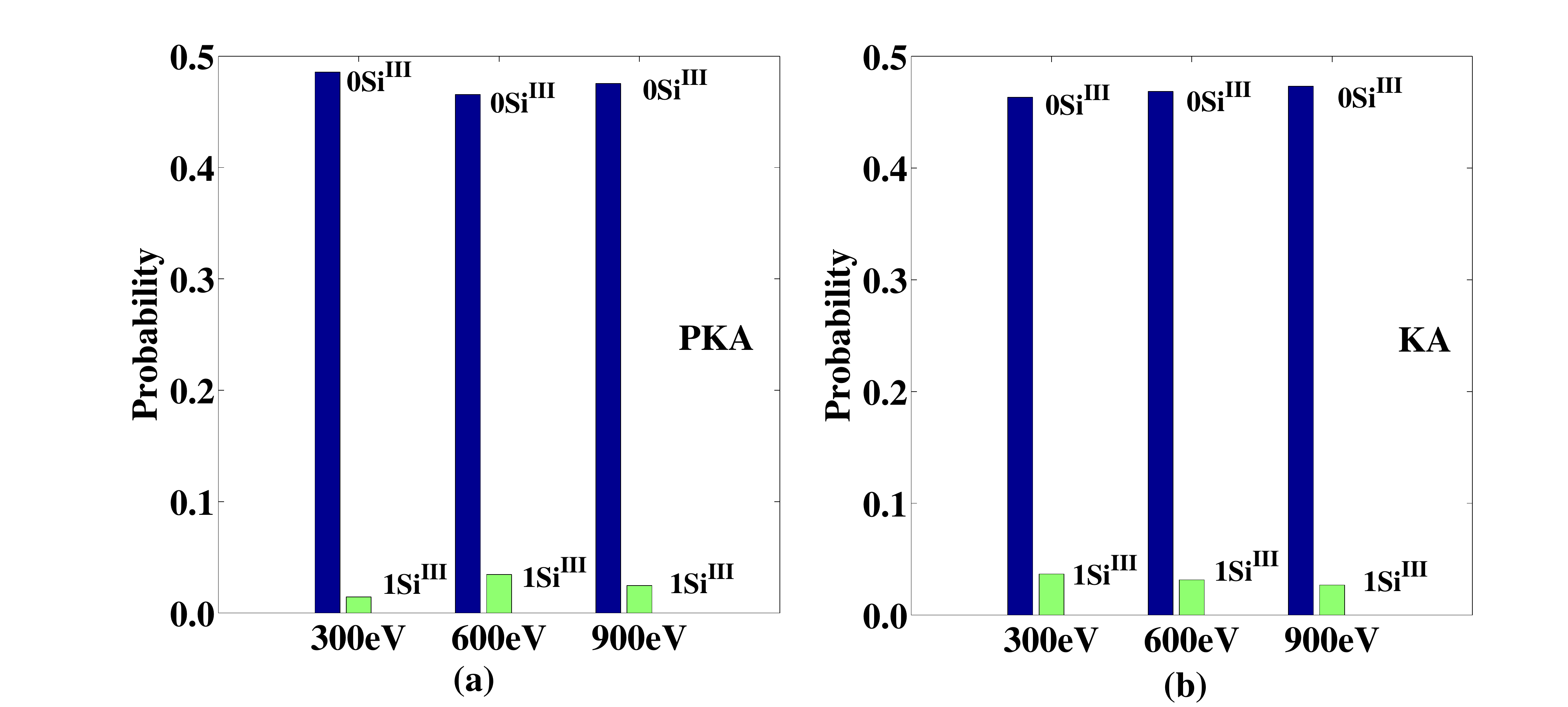}
\caption{\label{fig:2} Probabilities of creating zero or one three-coordinated Si (Si$^{\rm III}$) by (a) a O PKA, or (b) by all the knock-on atoms with displacement greater than 4 \AA\ generated by an O PKA. In any case, no simultaneous creation of two Si$^{\rm III}$ is found.}
\end{center}
\end{figure*}

In a cascade event (an example is shown in Fig. \ref{fig:1}, an atom absorbs the radiation energy and is accelerated (known as the Primary Knock-on Atom, or PKA). The high kinetic energy causes the PKA to substantially displace from its original position and, along the way, to collide with other atoms (Secondary Knock-on Atom, or SKA). In the case of an oxygen PKA, such a collisions do indeed create a vacancy at its original lattice site, immediately after displacement, leaving two three-coordinated Si atoms (Si$^{\rm III}$). However, we observe that a significant number of these Si$^{\rm III}$ disappear within a fairly short time (<10 ps). The analysis of 100 cascades for the three selected incident energies reveals that, on average, less than 7\%\ of the PKAs eventually generate oxygen vacancies (and Si$^{\rm III}$ species) at their original positions, as shown in Fig. \ref{fig:2}(a). In addition, for all the three incident energies, even in the case of Si$^{\rm III}$ formation, no oxygen PKA is observed to simultaneously generate and maintain both of the two Si$^{\rm III}$ species that are predicted by the ideal senario. The same feature is observed when tracking all the knock-on atoms (KA) with a displacement greater than 4 \AA\ (note that the inter-tetrahedron distance in $\rm \alpha$-quartz is 3.06 \AA), as shown in Fig. \ref{fig:2}(b).

\begin{figure*}
\scalebox{0.6}{\includegraphics*[width=\linewidth, keepaspectratio=true, draft=\ddst]{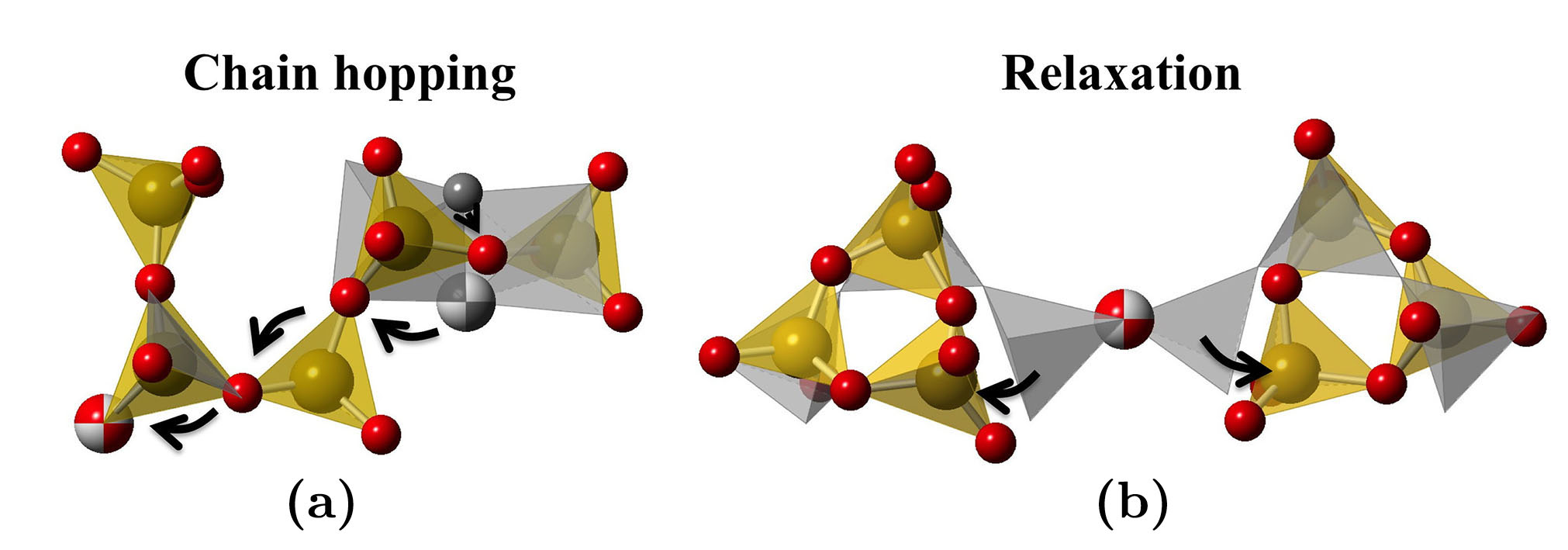}}
\caption{\label{fig:3} Mechanisms for oxygen vacancies removals. (a) shows a series of neighboring hops to achieve the recombination. (b) shows the relaxation of Si$^{\rm III}$ around the oxygen vacancy to form small Si--O rings and regain four-fold coordination. }
\end{figure*}

Further analyses of the cascade trajectories reveal three mechanisms for the annihilation of the oxygen vacancies following their initial creation. 
\begin{enumerate}
\item 	The first mechanism manifests by the recombination of the interstitial and vacancy. This does not necessarily require the knock-on atom to return to its original position, but rather it can be achieved by a chain of atoms hopping to their neighboring sites, as illustrated in  Fig. \ref{fig:3}(a). In our simulations, this has only been observed to occur when the displacement of the knock-on atom is smaller than 10 \AA.\ However, its occurrence at 300$\sim$900 eV incident energies is already rare so it should not be a dominant mechanism for oxygen vacancy annihilation under higher incident energies. 
\item 	The second mechanism, which is most common in our simulations, manifests by a local relaxation around the oxygen vacancy. After the initial formation of the two Si$^{\rm III}$, they can relax away from the vacancy and form small rings with the nearby silicon tetrahedra. This involves the formation of three member Si--O rings, as well as three-fold coordinated oxygen atoms, or tricluster oxygen, which are found to exist in some silicate glasses \cite{Bauchy2014}. An example is shown in  Fig. \ref{fig:3}(b), in which two three-member Si--O rings are formed by the two Si$^{\rm III}$ relax into an interstitial site in the quartz structure. 
\item The third mechanism, i.e., a replacement collision, involves a primary atom knocking out but at the same time replacing a SKA. This mechanism is found to be more common for Si KAs, as will be discussed later. 
\end{enumerate}

\begin{figure*}
\scalebox{1}{\includegraphics*[width=\linewidth, keepaspectratio=true, draft=\ddst]{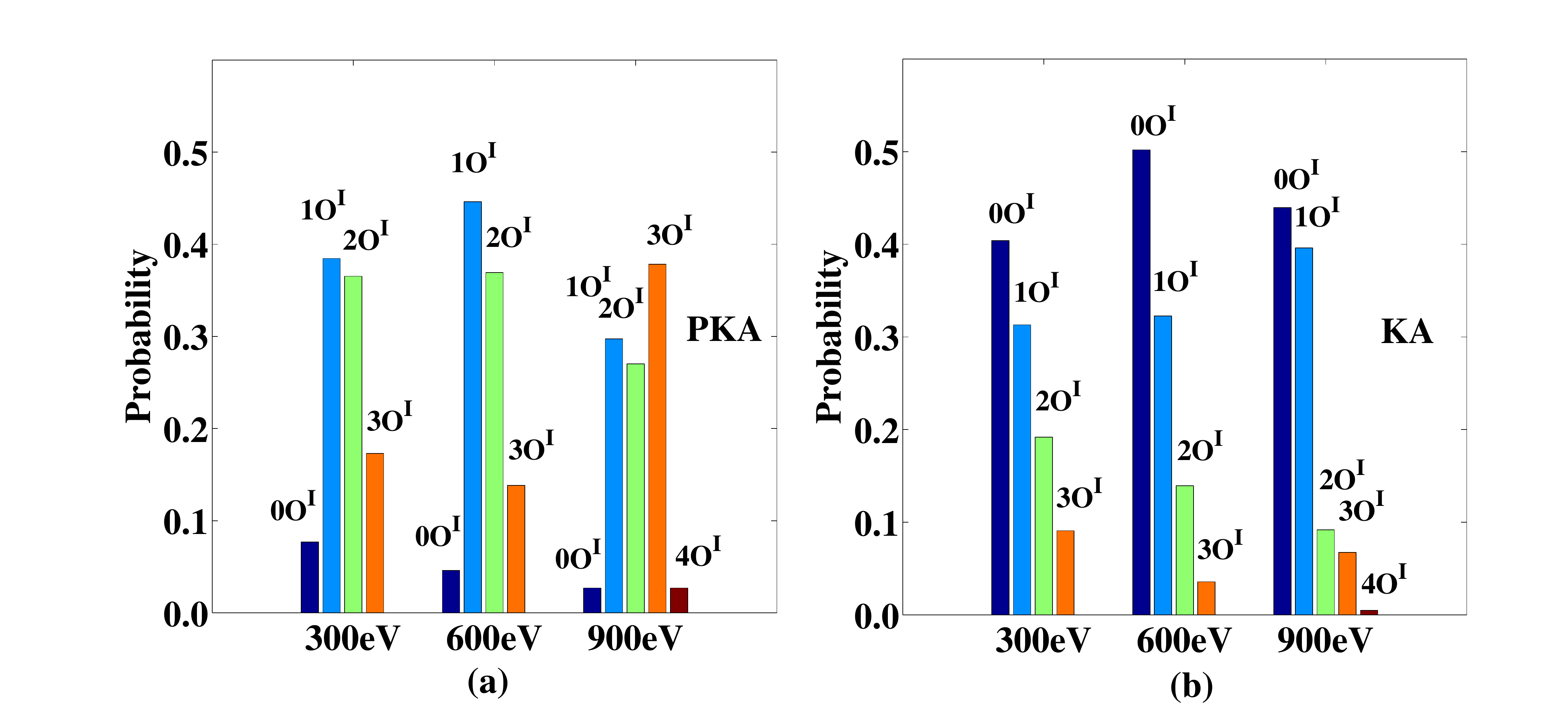}}
\caption{\label{fig:4} Probabilities of creating zero, one, two, three or four under-coordinated O (O$\rm ^{I}$) by (a) a Si PKA, or (b) by all the knock-on atoms with displacement greater than 4 \AA\ generated by a Si PKA.}
\end{figure*}

Other than oxygen vacancies, we note that other types of defects appear to form more readily during the cascade. As shown in Fig. \ref{fig:4}(a), a Si PKA almost always creates one or more under-coordinated oxygen atoms or, in other words, non-bridging oxygen atoms. For Si SKA, such probability decreases significantly due to the increased chance of replacement collisions and recombination with nearby interstitial defects, but the probability of forming non-bridging oxygens is still greater than 50\%\ for all three incident energies (see Fig. \ref{fig:4}(b)). At the end, the incident energy does not significantly impact the overall probability of creating dangling oxygen atoms by the knock-on atoms. It is worth noting that Si knock-on atoms can also lead to formations of small Si--O rings.

\begin{figure*}
\scalebox{1}{\includegraphics*[width=\linewidth, keepaspectratio=true, draft=\ddst]{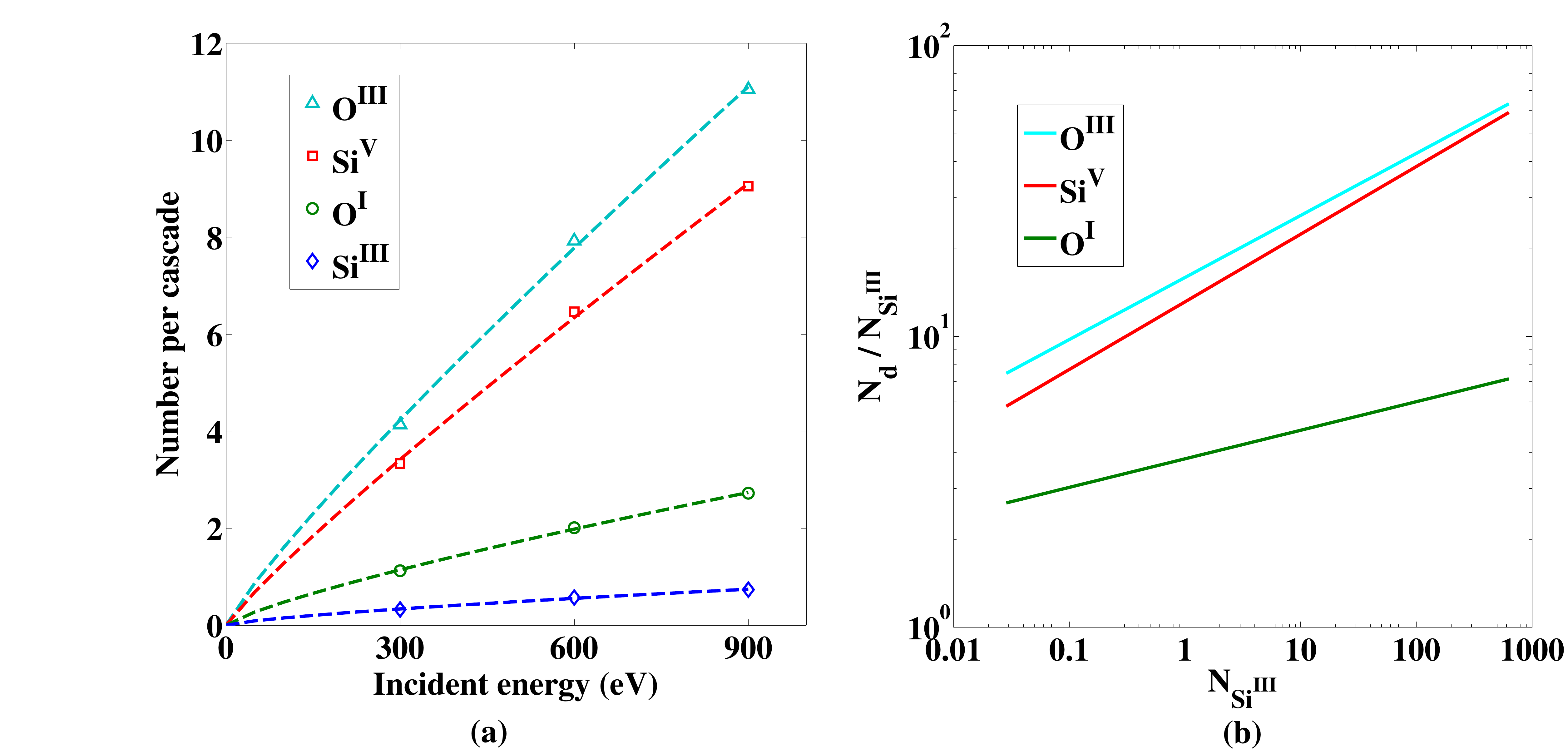}}
\caption{\label{fig:5} Average number of point defects created during one cascade is shown in (a). Data are fitted by power functions  $\rm N_{d}$= $AE^{m}$, represented by dashed lines. (b) shows the ratio between number of each defect type ($\rm N_{d}$) and number of Si$^{\rm III}$ ($\rm N_{Si^{III}}$), extrapolated to higher incident energies using the fitting results.}
\end{figure*}

Based on the results above, we note that different incident energies create similar types of point defects, i.e., under- and over-coordinated Si and O atoms. As illustrated in Fig. \ref{fig:1}, a knock-on atom usually creates under-coordinated defects at its original position. On the other hand, at the end of the cascade, displaced atoms usually remain in interstitial positions and form over-coordinated defects, such as tri-cluster oxygen and five-coordinated Si atoms. For a typical cone-shaped damage volume, at the end of the cascade, more over-coordinated defects are formed at the end of the cascade than under-coordinated defects around the initial knock-on atom. Indeed, the enumeration of all types of defects formed in a single cascade confirms such conclusion. As shown in Fig. \ref{fig:5}, oxygen vacancies (same quantity as Si$^{\rm III}$, as a knock-on atom dose not creates more than one oxygen vacancy) only constitute a small portion of the total defects for all incident energies. Also, we note that the ratio between different defect species is not a linear function of the incident energy. Indeed, the formation of over-coordination defects appears to be far more pronounced than that of the under-coordination ones. Therefore, the relative percentage of the oxygen vacancies, among all the point defects, decreases with the incident energy. As such, the $E^{\prime}$ center (charged oxygen vacancy) concentration alone may not be a proper indicator of radiation damage in quartz as illustrated in Fig. \ref{fig:5}(b) below.

\begin{center}
\begin{table}[h]
\caption{\label{tab:1} Fitting parameters for the energy dependence of the number of defects shown in Fig. \ref{fig:5}(a) (see text).}
\begin{tabular}{ccc}
\hline 
  & A (10$^{-2}$ /$eV^{m}$) & $m$ \\  
\hline 
Si$^{\rm III}$ & 0.539 & 0.72
 \\ 
O$^{\rm I}$
 & 
1.225
 & 0.80
 \\ 
Si$^{\rm V}$
 & 2.104
 & 0.89
 \\ 
O$^{\rm III}$
 & 2.820
 & 0.88
 \\ 
\hline 
\end{tabular} 
\end{table}
\end{center}

The above conclusions agree with experimental observations based on the energy loss during ion implantation (incident energy >30 keV), according to which up to 500 other defects could be created for every $E^{\prime}$ generated \cite{Devine1983,Devine1984}. However, this estimation may be imprecise as the nature of the defects, and their formation energies, was not precisely known. To provide a more accurate estimation, we establish an analytical relationship between the number of defects ($\rm N_{d}$) and the incident energy ($E$), which can then be used to extrapolate the data displayed in Fig. \ref{fig:5} to higher incident energies. Although the limited range of system sizes makes it difficult to determine the exact functional form of the relationship, it appears that they can be fitted by a power function of the form $\rm N_{d}$= $AE^{m}$. The corresponding parameters are listed in TABLE \ref{tab:1}. Hence, the calculated fraction of oxygen vacancies, with respect to the total amount of defects, decreases with the incident irradiation energy (3.6\%, 3.4\%, 3.1\%, at 300, 600 and 900 eV, incident energies respectively). As extrapolated to larger energies, this fraction would decrease to 2.2\%\ at 10 keV and 1.5\%\ at 100 keV. The knowledge of the fraction of each type of defect with respect to the incident energy, as presented here, can now be used to estimate the total number of defects formed, based on the measurement on the concentration of oxygen vacancies only. 

It must be noted that the point defects that are formed can cause substantial modifications to the Si--O network in quartz. For example, tri-cluster oxygen species are usually associated with edge-sharing Si tetrahedra, and three-member Si--O rings are created by the relaxation caused by vacancies. These network defects constitute a major portion of radiation damage in quartz. These aspects, together with more complex structural damage in radiated quartz, would warrant further studies.

\subsection{Threshold displacement energy}
\label{thresh_E}
\begin{figure*}
\includegraphics*[width=\linewidth, keepaspectratio=true, draft=\ddst]{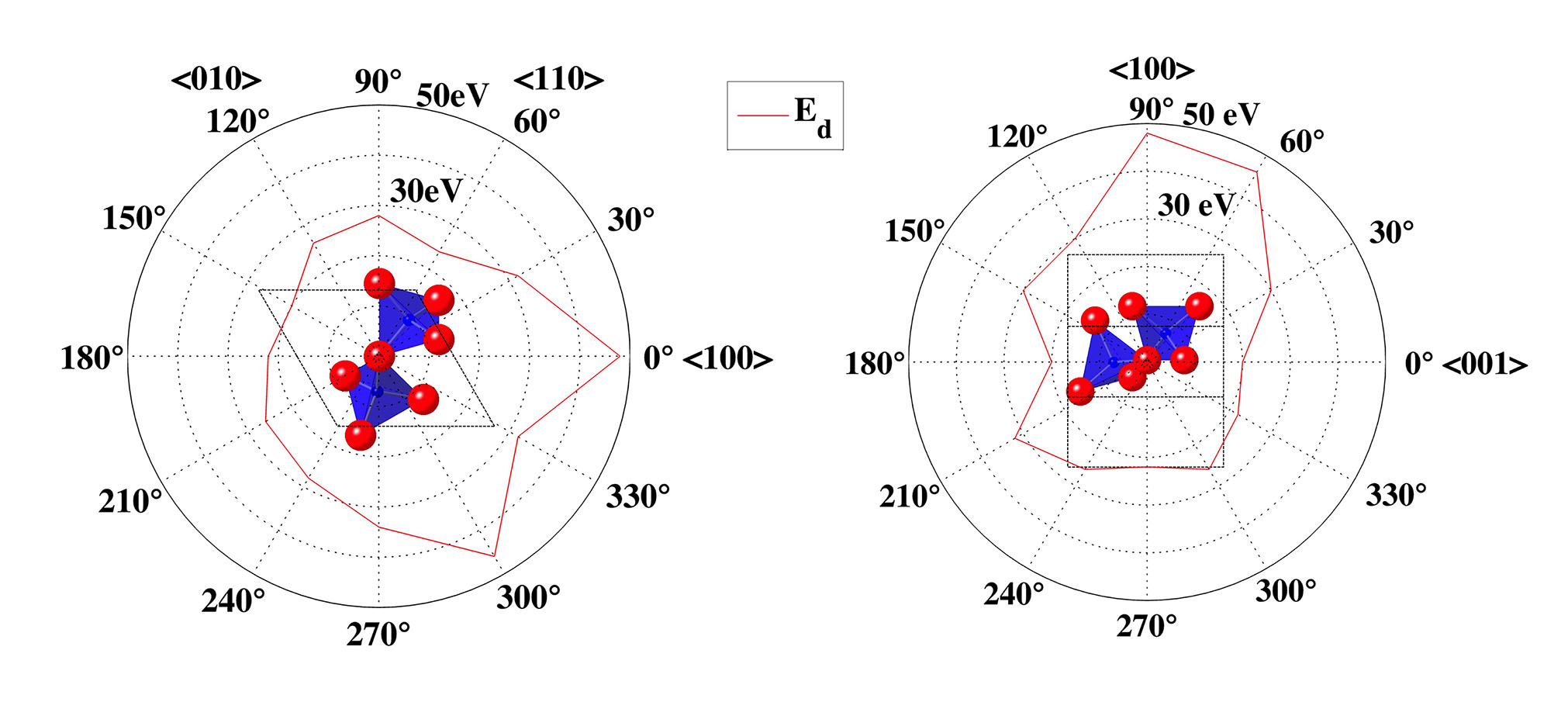}
\caption{\label{fig:6} Threshold displacement energy of oxygen along different directions in quartz crystal.}
\end{figure*}

\begin{figure*}
\includegraphics*[width=\linewidth, keepaspectratio=true, draft=\ddst]{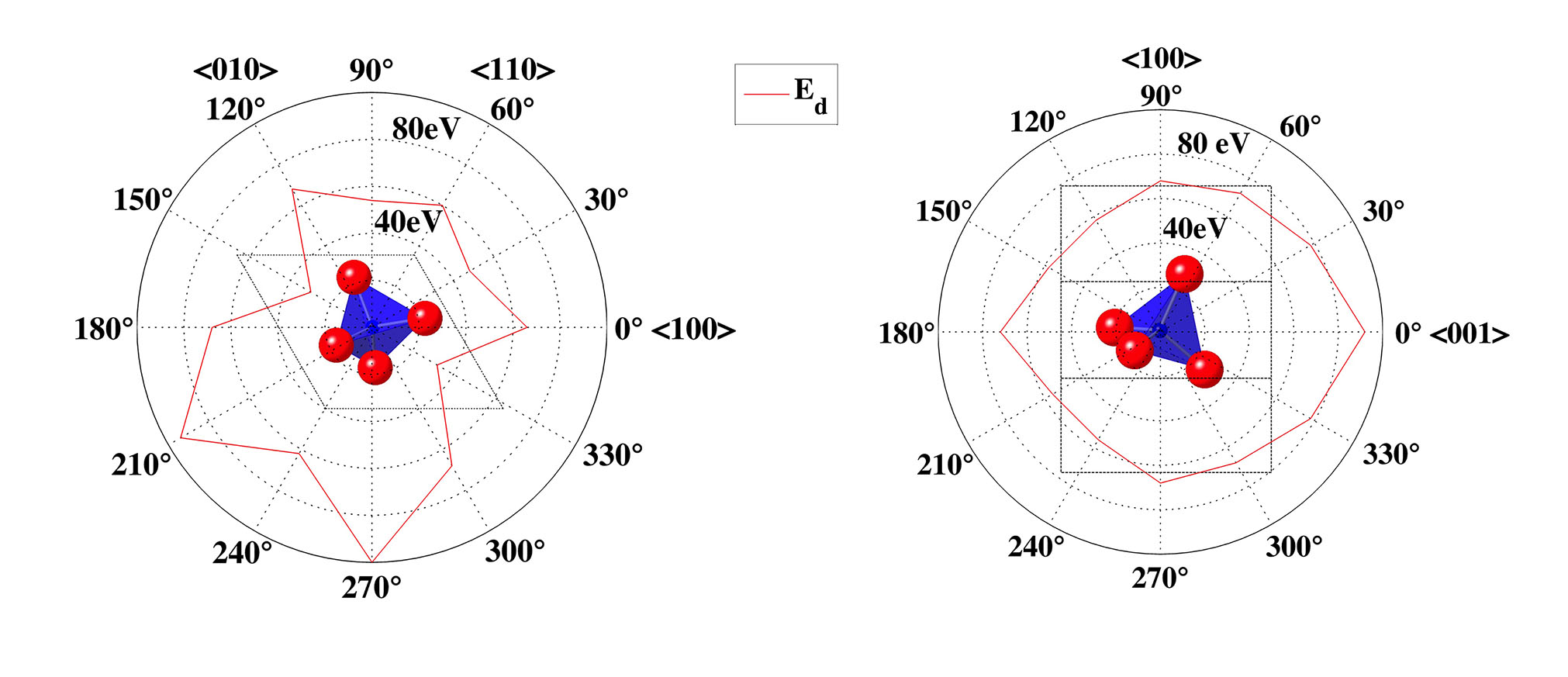}
\caption{\label{fig:7} Threshold displacement energy of silicon along different directions in quartz crystal.}
\end{figure*}

The threshold displacement energy is the energy needed to permanently displace an atom from its original lattice site. It is a critical input for the Monte Carlo method, which can be used to predict high-energy radiation damage for system sizes which are far too large to be handled by MD. This quantity is closely related to the atom types, the nature of bond and the crystal structure, and is difficult to determine experimentally. We utilize the MD technique to evaluate threshold displacement energies for both silicon and oxygen atoms. They are obtained by accelerating a given atom in a quartz supercell with increasing energy until a permanent displacement is observed. The simulation is performed along different directions, at 0K, and the results are presented in Fig. \ref{fig:6} and Fig. \ref{fig:7}. It should be noted however, that the temperature should also have an effect on the threshold displacement energy. In particular, we keep in mind that radiation exposure will induce local heating, which could affect the stability of the atoms and, thereby, their threshold displacement energy. Such effect is beyond the scope of the present study.

By plotting the threshold displacement energies for representative crystal planes, we observe a strong orientation dependence. In addition, the effect of the crystal symmetry is evident. Indeed, the minimum displacement energy is found to be 20 eV for oxygen along the <001> direction, and 30 eV for silicon along the <-110> direction among typical low-index orientations. Higher temperatures are likely to reduce the orientation dependence, resulting in a less pronounced orientation dependency. To predict the radiation damage in real applications, where the material is usually randomly orientated with respect to the incident radiation, threshold energies are averaged over 20 uniformly spaced spherical angles. We find the average displacement energies of 28.9 eV for oxygen and 70.5 eV for silicon. The threshold displacement energy for oxygen has a similar value to that found in fused silica by previous simulation studies using a different set of potentials \cite{Mota2004,Mota2009}. This is expected as the local atomic structure is similar in both materials. Nevertheless, the fact that results obtained from different potential sets are consistent with each other demonstrates the robustness of the present method and suggests that it can generally be applied to study radiation-induced damage in silicate minerals or glasses. The value, however, is considerably lower than the experimental value determined from optical measurements \cite{Arnold1965}, which may be related to the complex oxygen vacancy formation and annihilation process in irradiated quartz, as discussed in the previous section. We also note that the threshold energy for silicon determined from our simulations is around twice as large as that of oxygen, agreeing with the physical picture that twice as many bonds need to be broken to displace a silicon atom in comparison with an oxygen atom. Taken together, the validity of the present results, and their general nature suggests that MD simulation can indeed provide a realistic atomic picture of radiation-induced damage in quartz.

\section{Conclusion}
\label{sec:concloo}
Radiation damage in quartz is studied by detailed MD simulations. By analyzing hundreds of cascade trajectories, we find that the ballistic events cause a range of structural damage. At the end of the cascade, over-coordinated Si and O species are found to be the main types of point defects, and their abundance strongly depends on the incident energy. Some under-coordinated defects also exist around the knock-on atoms, but their quantities remain limited, due to recombination and replacement collisions. Oxygen vacancy defects are found to exist in a relatively low concentration, as compared with other defect types. In addition, this concentration decreases with the incident radiation energy. This suggests that the measurements of damage level based on $E^{\prime}$ center-related optical properties alone may not be a rigorous indicator of radiation damage. Structural damage also involves, in a considerable way, some modifications of the Si--O network, thereby creating small Si--O rings and edge-sharing Si tetrahedra. Such modifications could be the cause of radiation-induced amorphization, and should be investigated in further detail. We quantify the threshold displacement energy for both oxygen and silicon atoms from MD simulations. As expected, they are found to be strongly dependent on the crystal orientation. The average threshold displacement energy is determined to be 28.9 and 70.5 eV for oxygen and silicon, respectively. These quantitative results enable improved assessments of radiation damage in quartz.

\section*{Acknowledgements}
\label{sec:aknjm}
The authors acknowledge full financial support for this research provisioned by the University of California, Los Angeles (UCLA), Oak Ridge National Laboratory operated for the U.S. Department of Energy by UT-Battelle (Award \#\ 4000132990) and the National Science Foundation (CAREER Award \#\ 1253269). The authors acknowledge the support of these resources in making this research possible. The contents of this paper reflect the views and opinions of the authors, who are responsible for the accuracy of the datasets presented herein, and do not reflect the views and/or policies of the funding agencies, nor do the contents constitute a specification, standard or regulation.

\end{document}